\documentclass{article}
\usepackage{amsthm, amsmath, amsfonts, charter, mathrsfs, amssymb, bbding, authordate1-4,graphicx}

\begin{document}

\renewcommand{\qedsymbol}{$\blacksquare$}
\newtheorem{cor}{Corollary} 
\newtheorem{dfn}{Definition} 
\newtheorem{lem}{Lemma}
\newtheorem{prop}{Proposition}
\newtheorem{thm}{Theorem}

\title{Public Announcement Logic in Geometric Frameworks}
\author{Can Ba\c{s}kent}
\date{}
\maketitle

\begin{abstract}
In this paper we introduce public announcement logic in different geometric frameworks. First, we consider topological models, and then extend our discussion to a more expressive model, namely, subset space models. Furthermore, we prove the completeness of public announcement logic in those frameworks. Moreover, we apply our results to different issues: announcement stabilization, backward induction and persistence.
\end{abstract}

\section{Introduction}

Public announcement logic is a well-known and well-studied example of dynamic epistemic logics \cite{pla,dit}. Dynamic epistemic logics are set out to formalize knowledge and knowledge changes in usually multi-agent settings by defining and introducing different ways of updates and interaction. The contribution of public announcement logic (PAL, henceforth) to the field of knowledge representation is mostly due to its succinctness and clarity in reflecting the intuition as it does not increase the expressiveness of the basic epistemic logic. PAL updates the epistemic models by the announcements made by a truthful external agent. After the truthful announcement, the model is updated by eliminating the states that do not agree with the announcement. PAL has many applications in the fields of formal approaches to social interaction, dynamic logics, knowledge representation and updates \cite{bal3,bal2,ben10,ben4}. Extensive applications of PAL to different fields and frameworks has made PAL a rather familiar framework to many researchers. Moreover, virtually almost all applications of PAL make use of Kripke models for knowledge representation. However, as it is very well known, Kripke models are not the only representational tool for modal and epistemic logics.

In this work, we consider PAL in two different geometrical frameworks: topological models for modal logic and subset space logic. Topological models are not new to modal logics, indeed they are the first models for modal logic \cite{mck1,mck}. The past decades have witnessed a revival of academic interest towards the topological models for modal logics in many different frameworks \cite{aie2,ben2,ben1,bez2}. However, to the best of our knowledge, topological models have not been applied to dynamic epistemic logics. Yet, there have been some influential works on the notion of common knowledge in topological models which has motivated the current paper \cite{ben3}. In that work, it was shown that the different definitions of common knowledge diverge in topological models even though these definitions are equivalent in Kripke structures, based on Barwise's earlier investigation \cite{bar}. Nevertheless, the authors did not seem to take the next immediate step to discuss dynamic epistemologies in that framework. This is one of our goals in this paper: to apply topological reasoning to dynamic epistemological cases and present the immediate completeness results. The second framework that we discuss, subset space logic, is a rather weak yet expressive geometrical structure dispensing with the topological structure \cite{par9,par1}. Subset space logic has been introduced to reason about the topological notion of \emph{closeness} and the dynamic notion of \emph{effort} in epistemic situations. In this paper, we also define PAL in subset space logic with its axiomatization and present the completeness of PAL in subset space logics improving the results based on an earlier work \cite{bas2}.

There are several reasons that motivate this work. First, topological models can distinguish some epistemic properties that Kripke models cannot \cite{ben3}. This is perhaps not surprising as the topological semantics of the necessity modality has $\Sigma_2$ complexity, while Kripkean semantics offer $\Pi_1$ complexity for the same modality, and furthermore topologies deal with infinite cases in a rather special way\footnote{When we discuss the semantics of topological models, we will see the $\Sigma_2$ complexity of the aforementioned definition. Moreover, by definition, topologies do embrace infinite unions.}. Moreover, PAL update procedure is easily defined by using well-defined topological operations giving sufficient reasons to wonder what other different structures one may have in topological models.

The present paper is organized as follows. First, we introduce the geometrical frameworks that we need: topological spaces and subset spaces. Then, after a brief interlude on PAL, we give the axiomatizations of PAL in such spaces, and their completeness. The completeness proofs are rather immediate - which is usually the case in PAL systems. Then, we make some observations on PAL in geometric models. Our observations will be about the stabilization of updated models, backward induction in games and persistency.

\section{Geometric Models}

In this section, we will briefly recall the geometric models for some modal logics. What we mean by \emph{geometric} models is topological models and subset space logic models as they inherently are geometrical structures. We first start with topological models and their semantics, and then discuss subset space models.

\subsection{Topological Semantics for Modal Logic}

Topological interpretations for modal logic historically precede the relational semantics \cite{mck1,gold2}. Moreover, as we will observe very soon, topological semantics is arithmetically more complex than relational semantics: the prior is $\Sigma_2$ while the latter is $\Pi_1$. Now, let us start by introducing the definitions.

\begin{dfn} A topological space $\mathcal{S}= \langle S, \sigma \rangle$ is a structure with a set $S$ and a collection $\sigma$ of subsets of $S$ satisfying the following axioms:
\begin{enumerate}
\item The empty set and $S$ are in $\sigma$. 
\item The union of any collection of sets in $\sigma$ is also in $\sigma$.
\item The intersection of a finite collection of sets in $\sigma$ is also in $\sigma$.
\end{enumerate} \end{dfn}

The collection $\sigma$ is said to be a \emph{topology} on $S$. The elements of $S$ are called \emph{points} and the elements of $\sigma$ are called \emph{opens}. The complements of open sets are called \emph{closed} sets. Our main operator in topological spaces is called \emph{interior} operator $\mathbb{I}$ which returns the interior of a given set. The interior of a set is the largest open set contained in the given set. A topological model $\mathcal{M}$ is a triple $\langle S, \sigma, v \rangle$ where $\mathcal{S} = \langle S, \sigma \rangle$ is a topological space, and $v$ is a valuation function assigning propositional letters to subsets of $S$, i.e. $v: P \rightarrow \wp(S)$ for a countable set of propositional letters $P$. 

The basic modal language $\mathcal{L}$� has a countable set of proposition letters $P$, a truth constant� $\top$, the usual Boolean operators $\neg$ and $\wedge$, and a modal operator $\Box$. The dual of $\Box$ is denoted by $\Diamond$ and defined as $\Box \varphi \equiv \neg \Diamond \neg \varphi$. When we are in topological models, we will use the symbol $\mathsf{I}$ for $\Box$ after the \emph{interior} operator for intuitive reasons, and to prevent any future confusion. Likewise, we will use the symbol $\mathsf{C}$ for $\Diamond$. The notation $\mathcal{M}, s \models \varphi$ will read \emph{the point $s$ in the model $\mathcal{M}$ makes the formula $\varphi$ true}. We call the set of points that satisfy a given formula $\varphi$ in model $\mathcal{M}$ \emph{the extension} of $\varphi$, and denote as $(\varphi)^{\mathcal{M}}$. We will drop the superscript when the model we are in is obvious.

In topological models, the extension of a Boolean formula is obtained in the familiar sense. The extension of a modal formula in model $\mathcal{M}$, then, is given as follows $(\mathsf{I} \varphi)^{\mathcal{M}} = \mathbb{I}((\varphi)^{\mathcal{M}})$ - namely, the extension of $\mathsf{I} \varphi$ is the interior of the extension of $\varphi$. Now, based on this framework, the model theoretical semantics of modal logic in topological spaces is given as follows.

\begin{quote}\begin{tabular}{lll}
$\mathcal{M} , s \models p$ & iff & $s \in v(p)$ for $p \in P$ \\
$\mathcal{M} , s \models \neg \varphi$ & iff& $\mathcal{M}, s \not\models \varphi$  \\
$\mathcal{M} , s \models \varphi \wedge \psi$ & iff & $\mathcal{M} , s \models \varphi$ and $\mathcal{M} , s \models \psi$ \\
$\mathcal{M} , s \models \mathsf{I} \varphi$ & iff & $\exists U \in \sigma (s \in U \wedge \forall t \in U,~  \mathcal{M}, t \models \varphi)$ \\
$\mathcal{M} , s \models \mathsf{C} \varphi$ & iff& $\forall U \in \sigma (s \in U \rightarrow \exists t \in U,~ \mathcal{M}, t \models \varphi)$
\end{tabular}\end{quote}

A few words on the semantics are in order here. The necessity modality $\mathsf{I} \varphi$ says that there is an open set that contains the current state and the formula $\varphi$ is true everywhere in this set. Obviously, this is a rather complex statement, first, it requires us to determine the open set, and then check whether each point in this open set satisfies the given formula or not. On the other hand, the possibility  modality $\mathsf{C} \varphi$ manifests the idea that for every open set that includes the current state, there is point in the same set that satisfies $\varphi$. This is clearly reflected in the definition: in topological semantics, the definitions of modal satisfaction have the form $\exists \forall$ or $\forall \exists$. In Kripke models, as it is well-known, the form is either $\exists$ or $\forall$.

It is been shown by McKinsey and Tarski that the modal logic of topological spaces is S4 \cite{mck1}. Moreover, the logic of many other topological spaces has also been investigated \cite{aie2,cat,bez1,ben1,ben2}. Moreover, recently, the topological properties of paraconsistent systems have also been investigated \cite{bas3,mor}.

The proof theory of the topological models is as expected: we utilize modus ponens and necessitation. Basic modal logic is long to be known to be sound and complete with respect to the  well-known axiomatization of topological modal logic.

\subsection{Subset Space Logic}

Subset space logic (SSL, henceforth) was presented in early 90s as a bimodal logic to formalize reasoning about sets and points with an underlying motivation from epistemic logic \cite{par9}. One of the modal operators of SSL is intended to quantify \emph{over} the sets ($\Box$) whereas the other modal operator was intended to quantify \emph{in} the current set ($\mathsf{K}$). The underlying motivation for the introduction of these two modalities is to be able to speak about the notion of \emph{closeness}. In this context, $\mathsf{K}$ operator is intended for the knowledge operator (for one agent only, as SSL is originally presented for single-agent), and the $\Box$ modality is intended for the effort modality. Effort can correspond to various things: computation, observation, approximation - the procedures that can result in knowledge increase.

The language of subset space logic $\mathcal{L}_{S}$ has a countable set $P$ of propositional letters, a truth constant $\top$, the usual Boolean operators $\neg$ and $\wedge$, and two modal operators $\mathsf{K}$ and $\Box$. A subset space model is a triple $\mathcal{S} = \langle S, \sigma, v \rangle$ where $S$ is a non-empty set, $\sigma \subseteq \wp(S)$ is a collection of subsets (\emph{not} necessarily a topology), $v: P \rightarrow \wp(S)$ is a valuation function. Semantics of SSL, then is given inductively as follows.

\begin{quote}\begin{tabular}{llll}
$s, U \models p$ & iff & $s \in v(p)$ & \\
$s, U \models \varphi \wedge \psi$ & iff & $s, U \models \varphi$ & and~ ~ $s, U \models \psi$ \\
$s, U \models \neg \varphi$ & iff & $s, U \not\models \varphi$ & \\
$s, U \models \mathsf{K} \varphi$ & iff & $t, U \models \varphi$ & for all $t \in U$ \\
$s, U \models \Box \varphi$ & iff & $s, V \models \varphi$ & for all $V \in \sigma$ such that $s \in V \subseteq U$ \\ 
\end{tabular}\end{quote}
The duals of $\Box$ and $\mathsf{K}$ are $\Diamond$ and $\mathsf{L}$ respectively, and defined as usual. The tuple $(s, U)$ is called a \emph{neighborhood situation} if $U$ is a neighborhood of $s$, i.e. if $s \in U \in \sigma$. The axioms of SSL reflect the fact that the $\mathsf{K}$ modality is S5-like whereas the $\Box$ modality is S4-like. Moreover, we will need an additional axiom to state the interaction between those two modalities:  $\mathsf{K} \Box \varphi \rightarrow \Box \mathsf{K} \varphi$. Let us now give the complete set of axioms of SSL.
\begin{enumerate}
\item All the substitutional instances of the tautologies of the classical propositional logic
\item $(A \rightarrow \Box A) \wedge (\neg A \rightarrow \Box \neg A)$ for atomic sentence $A$  
\item $\mathsf{K} (\varphi \rightarrow \psi) \rightarrow (\mathsf{K} \varphi \rightarrow \mathsf{K} \psi)$
\item $\mathsf{K} \varphi \rightarrow (\varphi \wedge \mathsf{K} \mathsf{K} \varphi)$
\item $\mathsf{L} \varphi \rightarrow \mathsf{K} \mathsf{L} \varphi$
\item $\Box (\varphi \rightarrow \psi) \rightarrow (\Box \varphi \rightarrow \Box \psi)$
\item $\Box \varphi \rightarrow (\varphi \wedge \Box \Box \varphi)$
\item $\mathsf{K} \Box \varphi \rightarrow \Box \mathsf{K} \varphi$ 
\end{enumerate}

The rules of inference are as expected: modus ponens and necessitation for both modalities. Therefore, subset space logic is complete and decidable \cite{par9}.

Note that SSL is originally proposed as a single-agent system. There have been some attempts in the literature to suggest a multi-agent version of it, but to the best of our knowledge, there is no intuitive and clear presentation of a multi-agent version of SSL \cite{bas2}.

\subsection{Public Announcement Logic}{\label{Baskent-PAL}}

Public announcement logic is a way to represent changes in knowledge. The way PAL updates the epistemic states of the knower is by ``state-elimination''. A truthful announcement $\varphi$ is made, and consequently, the agents updates their epistemic states by eliminating the possible states where $\varphi$ is false \cite{pla,bal1,bal3,dit}.

Public announcement logic is typically interpreted on multi-modal (or multi-agent) Kripke structures \cite{pla}. Notationwise, the formula $[\varphi] \psi$ is intended to mean that \emph{after the public announcement of $\varphi$, $\psi$ holds}. As usual, $\mathsf{K}_{i}$ is the epistemic modality for the agent $i$. Likewise, $R_{i}$ is the epistemic accessibility relation for the agent $i$. The language of PAL will be that of multi-agent (multi-modal) epistemic logic with an additional public announcement operator $[*]$ where $*$ can be replaced with any well-formed formula in the language of basic epistemic logic. To see the semantics of PAL, take a model $\mathcal{M} = \langle W, \{ R \}_{i \in I}, V \rangle$ where $i$ denotes the agents and varies over a finite set $I$ . For atomic propositions, negations and conjunction the semantics is as usual. For modal operators, we have the following semantics.

\begin{quote}\begin{tabular}{ll}
$\mathcal{M}, w \models \mathsf{K}_{i} \varphi$ & iff $\mathcal{M}, v \models \varphi$ for each $v$ such that $(w, v) \in R_{i}$ \\
$\mathcal{M}, w \models [\varphi] \psi$ & iff $\mathcal{M}, w \models \varphi$ implies $\mathcal{M}| \varphi, w \models \psi$ 
\end{tabular}\end{quote}

Here, the updated model $\mathcal{M}|\varphi = \langle W', \{ R'_{i} \}_{i \in I}, V' \rangle$ is defined by restricting $\mathcal{M}$ to those states where $\varphi$ holds. Hence, $W' =  W \cap (\varphi)^{\mathcal{M}}$; $R_{i}' = R_{i} \cap (W' \times W')$, and finally $V'(p) = V(p) \cap W'$. The axiomatization of PAL is the axiomatization of S5$_n$ with additional axioms for dynamic modality. Hence, we give the set of axioms for PAL as follows.
\begin{enumerate}
\item All the substitutional instances of the tautologies of the classical propositional logic
\item $ \mathsf{K}_{i}(\varphi \rightarrow \psi) \rightarrow ( \mathsf{K}_{i} \varphi \rightarrow  \mathsf{K}_{i} \psi)$
\item $ \mathsf{K}_{i} \varphi \rightarrow \varphi$
\item $ \mathsf{K}_{i} \varphi \rightarrow  \mathsf{K}_{i}  \mathsf{K}_{i} \varphi$
\item $\neg  \mathsf{K}_{i} \varphi \rightarrow  \mathsf{K}_{i} \neg  \mathsf{K}_{i} \varphi$
\item $[\varphi] p \leftrightarrow (\varphi \rightarrow p)$ 
\item $[\varphi] \neg \psi \leftrightarrow (\varphi \rightarrow \neg [\varphi] \psi)$ 
\item $[\varphi] (\psi \wedge \chi) \leftrightarrow ([\varphi] \psi \wedge [\varphi] \chi)$ 
\item $[\varphi] \mathsf{K}_{i} \psi \leftrightarrow (\varphi \rightarrow \mathsf{K}_{i} [\varphi] \psi)$ 
\end{enumerate}
The additional rule of inference which we will need for announcement modality is called the \emph{announcement generalization} and is described as expected: From $\vdash \psi$, derive $\vdash [\varphi] \psi$.

PAL is complete and decidable. The completeness proof is quite straightforward. Once the soundness of the given axiomatization is proved, then it means that every complex formula in the language of PAL can be reduced to a formula in the basic language of (multi-agent) epistemic logic. Since S5 epistemic logic is long known to be complete, we immediately deduce the completeness of PAL. 

Notice again that in this section, we have defined PAL in Kripke structures by following the literature. In the next section, we will see how PAL is defined in geometrical models. We will start with SSL and proceed to topological models with some further observations.

\section{Subset Space PAL}

In SSL, we depend on neighborhood situations (which are tuples of the form $(s, U)$ for $s \in U \in \sigma$) instead of the epistemic accessibility relations. Therefore, if we want to adopt public announcement logic to the context of subset space logic, we first need to focus on the fact that the public announcements shrink the observation sets for each agent. 

Let us set a piece of notation. For a formula $\varphi$, recall that $(\varphi)^{\mathcal{S}}$ is the extension of $\varphi$ in the model $\mathcal{S} = \langle S, \sigma, v \rangle$. In SSL, $(\varphi)^{\mathcal{S}} = \{ (s, U) \in S \times \sigma : s \in U, (s, U) \models \varphi \}$. Define the projections $(\varphi)_1^\mathcal{S} : = \{ s : (s, U) \in (\varphi)^{\mathcal{S}} \text{ for some } U \ni s\}$, and $(\varphi)_2^\mathcal{S} : = \{ U : (s, U) \in (\varphi)^{\mathcal{S}} \text{ for some } s \in U \}$. We will drop the superscript when it is obvious.

Now, assume that we are in a subset space model $\mathcal{S} = \langle S, \sigma, v \rangle$. Then, after public announcement $\varphi$, we will move to another subset space model  $\mathcal{S}_{\varphi} = \langle S|\varphi, \sigma_{\varphi}, v_\varphi \rangle$ where $S|\varphi = (\varphi)_1$, and $\sigma_{\varphi}$ is the reduced collection of subsets after the public announcement $\varphi$, and $v_\varphi$ is the reduct of $v$ on $S|\varphi$. The crucial point is to construct $\sigma_{\varphi}$. As we need to get rid of the refutative states, we eliminate the points which do not satisfy $\varphi$ for each observation set $U$ in $\sigma$. We will disregard the empty set as no neighborhood situations can be formed with empty set. Hence, $\sigma_{\varphi} = \{ U_{\varphi} : U_{\varphi}= U \cap (\varphi)_2 \neq \emptyset, \text{ for each } U \in \sigma \}$. Alternatively, $\sigma_\varphi : = \{ U \cap (\varphi)_2 : U \in \sigma \} - \{ \emptyset \}$ \footnote{Thanks to the anonymous referee for pointing out this simple reformulation.}.

But then, how would the neighborhood situations be affected by the public announcements? Consider the neighborhood situation $(s, U)$ and the public announcement $\varphi$. Then the statement $s, U \models [\varphi] \psi$ will  mean that after the public announcement of $\varphi$, $\psi$ will hold in the neighborhood situation $(s, U_{\varphi})$. So, first we will remove the points in $U$ which refute $\varphi$, and then $\psi$ will hold in the updated set $U_{\varphi}$ which was obtained from the original set $U$. Then the corresponding semantics can be suggested as follows:
$$s, U \models [\varphi] \psi \text{ iff } s, U \models \varphi \text{ implies } s, U_{\varphi} \models \psi $$
Before checking whether this semantics satisfies the axioms of public announcement logic, let us give the language and semantics of the topologic PAL. The language of the topologic public announcement logic interpreted in subset spaces is given as follows:
\begin{center}
$ p ~|~ \bot ~|~ \neg \varphi ~|~ \varphi \wedge \psi ~|~ \Box \varphi ~|~ \mathsf{K} \varphi ~|~ [\varphi] \psi $
\end{center}

Now, let us consider the soundness of the axioms of basic PAL that we discussed earlier in Section~\ref{Baskent-PAL}. We prove that those axioms are sound in SSL.

\begin{thm} Axioms of the basic PAL are sound in subset space logic. \end{thm}
\begin{proof}
As the atomic propositions do not depend on the neighborhood, the first axiom is satisfied by the subset space semantics of public announcement modality. To see this, assume $s, U \models[\varphi] p$. So, by the semantics $s, U \models \varphi$ implies $s, U_{\varphi} \models p$. So, $s \in v(p)$. So for any set $V$ where $s \in V$, we have $s, V \models p$. Hence, $s, U \models \varphi$ implies $s, U \models p$, that is $ s, U \models \varphi \rightarrow p$. Conversely, assume $ s, U \models \varphi \rightarrow p$. So, $s, U \models \varphi $ implies $s \in v(p)$. As $s, U \models \varphi$, $s$ will lie in $U_{\varphi}$, thus $(s, U_{\varphi})$ will be a neighborhood situation. Thus, $s, U_{\varphi} \models p$. Then, we conclude $s, U \models [\varphi] p$.

The axioms for negation and conjunction are also straightforward formula manipulations and hence skipped.

The important reduction axiom is the knowledge announcement axiom. Assume, $s, U \models [\varphi] \mathsf{K} \psi$. Suppose further that $s, U \models \varphi$. Then we have the following.

\begin{quote} \begin{tabular}{lll}
$s, U \models [\varphi] \mathsf{K} \psi$ & iff & $s, U_{\varphi} \models \mathsf{K} \psi$ \\
& iff & for each $t_{\varphi} \in U_{\varphi}$, we have $t_{\varphi}, U_{\varphi} \models \psi$ \\
& iff & for each $t \in U$, $t, U \models \varphi$ \\ && implies $t, U \models [\varphi] \psi$ \\
& iff & $s, U \models \mathsf{K}(\varphi \rightarrow [\varphi] \psi)$ \\
& iff & $s, U \models \mathsf{K} [\varphi] \psi$
\end{tabular} \end{quote} 

Thence, the above axioms are sound for the subset space semantics of public announcement logic.
\end{proof}

Now, recall that SSL has an indispensable modal operator $\Box$. One can wonder whether we can have a reduction axiom for it as well. We start by considering the statement $[\varphi] \Box \psi \leftrightarrow (\varphi \rightarrow \Box [\varphi] \psi)$. Assume, $s, U \models [\varphi] \Box \psi$. Suppose further that $s, U \models \varphi$. Then,  
we deduce the following.
\begin{quote} \begin{tabular}{lll}
$s, U \models [\varphi] \Box \psi$ & iff & $s, U_{\varphi} \models \Box \psi$ \\
& iff & for each $V_{\varphi} \subseteq U_{\varphi}$ we have $s, V_{\varphi} \models \psi$ \\
& iff & for each $V \subseteq U$, $s, V \models \varphi$ \\ && implies $s, V \models [\varphi] \psi$ \\
& iff & $s, U \models \Box (\varphi \rightarrow [\varphi] \psi)$ \\
& iff & $s, U \models \Box [\varphi] \psi$
\end{tabular} \end{quote} 
Now, it is easy to see that the following axiomatize the SSL-PAL together with the axiomatization of SSL:
\begin{enumerate}
\item $[\varphi] p \leftrightarrow (\varphi \rightarrow p)$
\item $[\varphi] \neg \psi \leftrightarrow (\varphi \rightarrow \neg [\varphi] \psi)$ 
\item $[\varphi] (\psi \wedge \chi) \leftrightarrow ([\varphi] \psi \wedge [\varphi] \chi)$
\item $[\varphi] \mathsf{K} \psi \leftrightarrow (\varphi \rightarrow \mathsf{K} [\varphi] \psi)$
\item $[\varphi] \Box \psi \leftrightarrow (\varphi \rightarrow \Box [\varphi] \psi)$ 
\end{enumerate} 
Referring to the above discussions, the completeness of subset space PAL follows easily.

\begin{thm}
PAL in subset space models is complete with respect to the axiom system given above.
\end{thm}

\begin{proof}
By reduction axioms we can reduce each formula in the language of topologic PAL to a formula in the language of SSL. As SSL is complete, so is PAL in subset space models.
\end{proof}

By the same idea, we can import the decidability result.
\begin{thm} PAL in subset space models is decidable.
\end{thm}

\section{Topological PAL}

\subsection{Single Agent Topological PAL}

We can use the similar ideas to give an account of PAL in topological spaces. Let $\mathcal{T} = \langle T, \tau, v \rangle$ be a topological model and $\varphi$ be a public announcement. We now need to obtain the topological model $\mathcal{T}_{\varphi}$ which is the updated model after the announcement. Define $\mathcal{T}_{\varphi} = \langle T_{\varphi}, \tau_{\varphi}, v_{\varphi} \rangle$ where $T_{\varphi} = T \cap (\varphi)$, $\tau_{\varphi} = \{ O \cap T_{\varphi} : O \in \tau \}$ and $v_{\varphi} = v \cap T_{\varphi}$. We now need to verify that $\tau_{\varphi}$ is a topology, indeed the induced topology. For the sake of the completeness of our arguments in this paper, let us give the immediate proof here.

\begin{prop} If $\tau$ is a topology, then $\tau_{\varphi} = \{ O \cap T_{\varphi} : O \in \tau \}$ is a topology as well.
\end{prop}

\begin{proof} Clearly, the empty set is in $\tau_{\varphi}$ as $\tau$ is a topology. As $\tau$ is a topology on $T$, we have $T \in \tau$. Thus, $T \cap T_{\varphi}$, namely $T_{\varphi}$, is in $\tau_{\varphi}$. Consider $\bigcup_{i}^{\infty} U_{i}$ where $U_{i} \in \tau_{\varphi}$. For each $i$, we have $U_{i} = O_{i} \cap T_{\varphi}$ for some $O_{i} \in \tau$. Thus, $\bigcup_{i}^{\infty} U_{i} = T_{\varphi} \cap \bigcup_{i}^{\infty} O_{i}$. Since $\tau$ is a topology, $\bigcup_{i}^{\infty} O_{i} \in \tau$. Thus, $T_{\varphi} \cap \bigcup_{i}^{\infty} O_{i} \in \tau_{\varphi}$ yielding the fact that $\bigcup_{i}^{\infty} U_{i} \in \tau_{\varphi}$. Similarly, consider $\bigcap_{i}^{n} U_{i}$ where $U_{i} \in \tau_{\varphi}$ for some $n < \omega$. Since $U_{i} = O_{i} \cap T_{\varphi}$ for some $O_{i} \in \tau$, we similarly observe that $\bigcap_{i}^{n} U_{i} = \bigcap_{i}^{n} (O_{i} \cap T_{\varphi}) = T_{\varphi} \cap \bigcap_{i}^{n} O_{i}$. Since $\tau$ is a topology, $\bigcap_{i}^{n} O_{i} \in \tau$, thus, $\bigcap_{i}^{n} U_{i} \in \tau_{\varphi}$.
\end{proof}

It is important to notice here that only modal formulas necessarily yield open or closed extensions. The extension of Booleans, then, may or may not be a topological set as it solely depends on the model.

Now, when we restrict the carrier set of the topology to a subset of it, we still get a topology immediately and easily. Based on this simple observation, we can give a semantics for the public announcements in topological models.
$$\mathcal{T}, s \models [\varphi] \psi \text{ iff } \mathcal{T}, s \models \varphi \text{ implies }\mathcal{T}_{\varphi}, s \models \psi$$

In a similar fashion, we can expect that the reduction axioms work in topological spaces. The reduction axioms for atoms and Booleans are quite straight-forward. So, consider the reduction axiom for the interior modality given as follows: $[\varphi] \mathsf{I} \psi \leftrightarrow (\varphi \rightarrow \mathsf{I} [\varphi] \psi)$. 

Let $\mathcal{T}, s \models [\varphi] \mathsf{I} \psi$ which, by definition means $\mathcal{T}, s \models \varphi$  implies $\mathcal{T}_{\varphi}, s \models  \mathsf{I} \psi$. If we spell out the topological interior modality, we get $\exists U_{\varphi} \ni s \in \tau_{\varphi}$ s.t. $\forall t \in U_{\varphi}, \mathcal{T}_{\varphi}, t \models \psi$. By definition, since $U_{\varphi} \in \tau_{\varphi}$, it means that there is an open $U \in \tau$ such that $U_{\varphi}= U \cap (\varphi)$. Under the assumption that $\mathcal{T}, s \models \varphi$, we observe that $\exists U \ni s \in \tau$ (as we just constructed it), such that after the announcement $\varphi$, the non-eliminated points in $U$ (namely, the ones in $U_{\varphi}$) will satisfy $\psi$. Thus, we get $\mathcal{T}, s \models \varphi \rightarrow \mathsf{I} [\varphi] \psi$.

The other direction is very similar and hence we leave it to the reader. Therefore, the reduction axioms for PAL in topological spaces are given as follows.
\begin{enumerate}
\item $[\varphi] p \leftrightarrow (\varphi \rightarrow p)$ 
\item $[\varphi] \neg \psi \leftrightarrow (\varphi \rightarrow \neg [\varphi] \psi)$
\item $[\varphi] (\psi \wedge \chi) \leftrightarrow ([\varphi] \psi \wedge [\varphi] \chi)$
\item $[\varphi] \mathsf{I} \psi \leftrightarrow (\varphi \rightarrow \mathsf{I} [\varphi] \psi)$
\end{enumerate}
As a result, all the complex formulas involving the PAL operator can be reduced to a simpler one. This algorithm directly shows the completeness of PAL in topological spaces by reducing each formula in the language of topological PAL to the language of basic topological modal logic. Thus, the result follows.

\begin{thm} PAL in topological spaces is complete with respect to the axiomatization given.
\end{thm}

By the same idea, we can import the decidability result.
\begin{thm} PAL in topological models is decidable.
\end{thm}

\subsection{Product Topological PAL}

There are variety of ways to merge given topological models to express the epistemic interaction between them: products, sums, fusions etc \cite{gabb}. In this section, we focus on one of such methods, product topologies, and discuss how public announcements are defined in them. Product topological frameworks for multi-agent epistemic logics have already been discussed in the literature widely \cite{ben1,ben3}. Therefore, our treatment of the subject will be based on these works. Based on this basic formalism, we will then introduce public announcement logic. The idea is quite straight-forward. We are given two topologies (possibly with different spaces) with a modal (epistemic, doxastic etc) model on them. Then, by the standard techniques in the literature, we merge them. After that, we discuss how public announcements work in this unified structure.

Let $\mathcal{T} = \langle T, \tau \rangle$ and $\mathcal{T}' = \langle T', \tau' \rangle$ be two given  topological spaces. Now, we introduce some definitions. Let $X \subseteq T \times T'$. We call $X$ horizontally open (\emph{h-open}) if for any $(x, y) \in X$, there is a $U \in \tau$ such that $x \in U$, and $U \times \{ y\} \subseteq X$. In a similar fashion, we call $X$ vertically open (\emph{v-open}) if or any $(x, y) \in X$, there is a $U' \in \tau'$ such that $y \in U'$, and $\{ x \} \times U' \subseteq X$. These notions can be seen as one dimensional projections of openness and closure that we will need soon.

Now, given two topological spaces $\mathcal{T} = \langle T, \tau \rangle$ and $\mathcal{T}' = \langle T', \tau' \rangle$, let us associate two modal operators $\mathsf{I}$ and $\mathsf{I}'$ respectively to these models. Then, we can obtain a product topology in a language with the two aforementioned modalities. The product model, then, is of the form $\langle T \times T', \tau, \tau' \rangle$. Therefore, we consider the cross product $\times$ as a way to represent model interaction among epistemic agents which gives us a model with two-dimensional space, and two topologies. 

The semantics of those modalities are given as follows.
\begin{quote} \begin{tabular}{lll}
$(x, y) \models \mathsf{I} \varphi$ & iff & $\exists U \in \tau$, $x \in U$ and $ \forall u \in U$, $(u, y) \models \varphi$ \\
$(x, y) \models \mathsf{I}' \varphi$ & iff & $\exists U' \in \tau'$, $y \in U'$ and $ \forall u' \in U'$, $(x, u') \models \varphi$
\end{tabular}\end{quote}

Here, given a tuple $(x, y)$, the modality $\mathsf{I}$ ranges over the first component while the modality $\mathsf{I}'$ ranges over the second. In other words, we localize the product with respect to the given original topologies.

It has been shown that the fusion logic S4$\oplus$S4 is complete with respect to products of arbitrary topological spaces \cite{ben3}. Then, the question is this: How would a state elimination based dynamic epistemic paradigm work in product topologies? 

Now, step by step, we will present how to define public announcements in this framework. The difficulty lies in the fact that when we take the product of the given topological models, we increase the dimension of the space. Then, the intuition behind defining public announcements should follow the same idea: the announcement will update the product topology in all dimensions.

Let us now be a bit more precise. Before we start, note that here we focus on the product of two topologies representing the interaction between two agents with different spaces and topologies, but it can easily be generalized to $n$-agents.
The language of product topological PAL is given as follows.
$$ p ~|~ \neg \varphi ~|~ \varphi \wedge \varphi ~|~ \mathsf{K}_1 \varphi ~|~ \mathsf{K}_2 \varphi ~|~ [\varphi] \varphi$$

For given two topological models $\mathcal{T} = \langle T, \tau, v \rangle$ and $\mathcal{T}' = \langle T', \tau', v \rangle$, the product topological model $M = \langle T \times T', \tau, \tau', v \rangle$ has the following semantics.  
\begin{quote} \begin{tabular}{lll}
$M, (x, y) \models \mathsf{K}_1 \varphi$ & iff & $\exists U \in \tau$, $x \in U$ and $ \forall u \in U$, $(u, y) \models \varphi$ \\
$M, (x, y) \models \mathsf{K}_2 \varphi$ & iff & $\exists U' \in \tau'$, $y \in U'$ and $ \forall u' \in U'$, $(x, u') \models \varphi$ \\
$M, (x, y) \models [\varphi] \psi$ & iff & $M, (x, y) \models \varphi$ implies $M_\varphi, (x, y) \models \psi$
\end{tabular}\end{quote}
where $M_\varphi = \langle T_\varphi \times T'_\varphi, \tau_\varphi, \tau'_\varphi, v_\varphi \rangle$ is the updated model. We define all  $T_\varphi$, $T'_\varphi$, $\tau_\varphi$, $\tau'_\varphi$, and $v_\varphi$ as before. Therefore, the following axioms axiomatize the product topological PAL together with the axioms of S4$\oplus$S4.
\begin{enumerate}
\item $[\varphi] p \leftrightarrow (\varphi \rightarrow p)$
\item $[\varphi] \neg \psi \leftrightarrow (\varphi \rightarrow \neg [\varphi] \psi)$ 
\item $[\varphi] (\psi \wedge \chi) \leftrightarrow ([\varphi] \psi \wedge [\varphi] \chi)$
\item $[\varphi] \mathsf{K}_i \psi \leftrightarrow (\varphi \rightarrow \mathsf{K}_i [\varphi] \psi)$
\end{enumerate} 

\begin{thm} Product topological PAL is complete and decidable with respect to the given axiomatization.
\end{thm}

\begin{proof} Proof of both completeness and decidability is by reduction, and similar to the ones presented before. Thus, we leave the details to the reader.
\end{proof}

\section{Applications}

Now, we can briefly apply the previous discussions to some issues in PAL, foundational game theory and SSL. The purpose of such applications is to give the reader a sense how topological frameworks might affect the aforementioned issues, and in general how dynamic epistemic situations can be represented topologically.

\subsection{Announcement Stabilization}

Muddy Children presents an interesting case for PAL \cite{fag1}. In this game, we assume that a group of children were playing outside in the mud. Then, their father calls them in. Children came back in, and gather around the father in such a way that every children sees all the others, and the father sees them all. We also assume that there is no mirror in the room, so the children cannot see themselves. Since they were playing in the mud, some got dirty with mud on their forehead. Father then announces that ``At least one of you has mud on his or her forehead". If no child steps forward saying that ``Yes, I do have mud on my forehead" communicating the fact that she learned it from the announcement, the father keeps repeating the very same announcement \cite{dit}.

In that game, the model representing the epistemics of the group (see the Figure) gets updated after each children says that she does not know if she had mud on her forehead. The model keeps updated until the announcement is negated, and then becomes common knowledge \cite{ben18}. Therefore, after each update, we get smaller and smaller models up until the moment that the model gets stabilized in the sense that the same announcement does not update the model any longer.

\begin{figure}[ht]
\includegraphics[width=\textwidth]{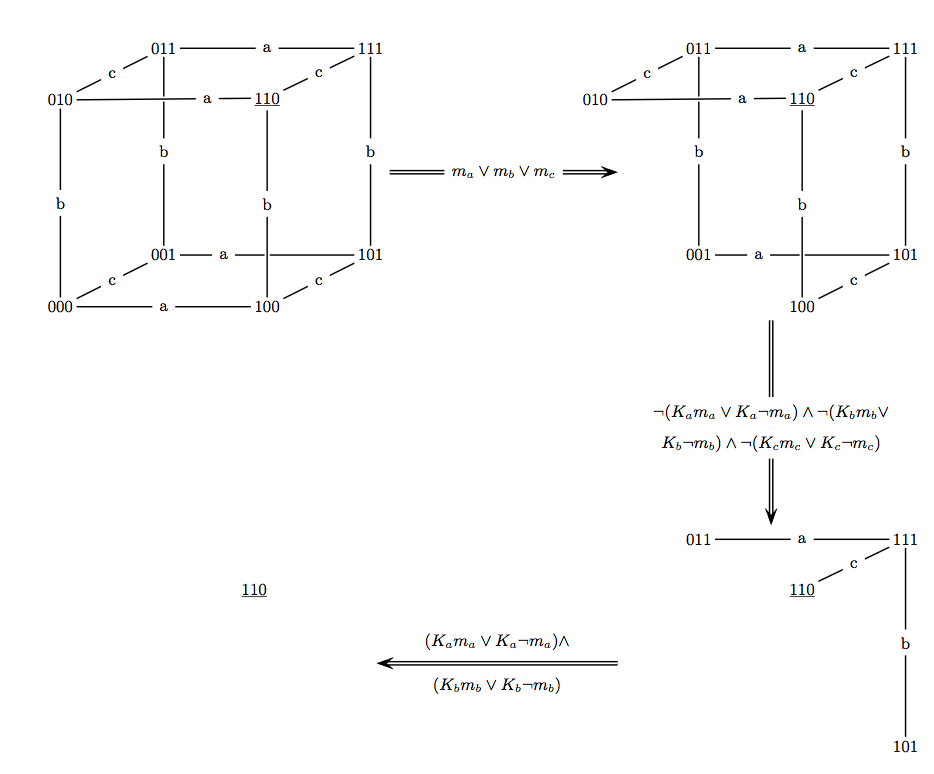}
\caption{A model for muddy children played with 3 children $a, b, c$ taken from van Ditmarsch \emph{et al}. The state $n_{a}n_{b}n_{c}$ for $n_{a}, n_{b}, n_{c} \in \{ 0, 1\}$ represent that child $i$ has mud on her forehead iff $n_i = 1$ for $i \in \{ a, b, c\}$. The proposition $m_i$ means that the child $i \in \{ a, b, c \}$ has mud on her forehead.  The current state is underlined.}
\end{figure}

As van Benthem pointed out, this is closely related to several issues in modal and epistemic logics \cite{ben18}. First, PAL behaves like a fixed-point operator where the fixed point is the model which is stabilized. Second, there seems to be a close relation between game theoretical strategy eliminations, and solution methods based on such approaches. Therefore, it is rather important to analyze announcement stabilization. Here, we will approach the issue from a topological angle.

For a model $M$ and a formula $\varphi$, we define the announcement limit $\lim_{\varphi} M$ as the first model which is reached by successive announcements of $\varphi$ that no longer changes after the last announcement is made. Announcement limits exist in both finite and infinite models \cite{ben14}. For instance, for any model $M$, $\lim_{p}M = M|p$ for propositional variable $p$. Therefore, the limit model is the first updated model when the announcement is a ground Boolean formula. In muddy children, the announcement shrinks the model step by step, round by round \cite{ben18}. However, sometimes in dialogue games it may take \emph{too long} to solve such puzzles until the model gets stabilized as shown by Parikh \cite{par10}. Similarly, even Zermelo considered similar approaches in early 20. century to understand as to how long it takes for the game to stabilize \cite{sch}.

Similar to the discussions of the aforementioned authors, we now analyze how the models stabilize in topological PAL. We know that topological models do present some differences in epistemic logical structures. For instance, in topological models, the stabilization of the fixed-point definition\footnote{Formula $\varphi$ is common knowledge among two-agents 1 and 2 $C_{1,2} \varphi$ is represented with the (largest) fixed-point definition as follows: $C_{1,2} \varphi : = \nu p. \varphi \wedge \mathsf{K}_1 p \wedge \mathsf{K}_2 p$ where $\mathsf{K}_i$, for $i=1,2$ is the familiar knowledge operator \cite{bar}.} version of common knowledge may occur later than ordinal stage $\omega$. However, it stabilizes in $\leq \omega$ steps in Kripke models \cite{ben3}. 

We also know that there are two possibilities for the limit models. Either it is empty or nonempty. If it is empty, it means that the negation of the announcement has become common knowledge, thus the announcement refuted itself. On the other hand, if the limit model is not empty, it means that the announcement has become common knowledge \cite{ben14}.

\begin{thm}{\label{baskent-limit}}
For some formula $\varphi$ and some topological model $M$, it may take more than $\omega$ stage to reach the limit model $\lim_{\varphi} M$.
\end{thm}

\begin{proof} 
The proof is rather immediate for those familiar with the literature. So, we just mention the basic idea here.

First, note that it was shown that in multi-agent topological models, stabilization of common knowledge with fixed-point definition may occur later than $\omega$ stage. However, in Kripke models it occurs before $\omega$ stage \cite{ben3}.

Also note that it was also shown that if the limit model is not empty, the announcement has become common knowledge \cite{ben14}.

Therefore, combining these two observations, we conclude that in some topological models with non-empty limit models, the number of stage for the announcement to be common knowledge may take more than $\omega$ steps.
\end{proof}

Even if the stabilization takes longer, we can still obtain stable models by taking intersections at the limit ordinals as a general rule \cite{ben14}. Therefore, we guarantee that the update procedure will terminate. Thus, the following result is now self-evident.

\begin{thm} Limit models exist in topological models. \end{thm}

Yet another property of topological models is the fact that the topologies are \emph{not} closed under arbitrary intersection. Then, one can ask the following question: ``How does PAL work in infinite-conjunction announcements?'' The following example illustrates that point. Take the real closed interval $[-1, 1]$ with the usual Euclidean topology. For each $n \in \omega$, define the valuation for propositions as such $v(p_n) = [-1/n, 1/n]$. Therefore, $p_1$ holds in the entire space $[-1, 1]$, while $p_2$ holds in $[-1/2, 1/2]$. Consider now the announcements $\Box \bigwedge_{n \in \omega} p_n$ and $\bigwedge_{n \in \omega} \Box p_n$. The former formula is true in the interior $\mathbb{I}(\bigcap_{n \in \omega} p_n)$ which is equal to empty set while the latter one is true in the intersection $\bigcap_{n \in \omega} \mathbb{I}(p_n)$ which is equal to the singleton $\{ 0 \}$. Then, clearly these updates will yield the same models in Kripke models. But, in topological models, as the extensions of two formula differ, updated models will clearly differ, too.

\subsection{Backward Induction}

The fact that limit models can be attained in more than $\omega$ steps can create some problems in games. Consider the backward induction solution where players trace back their moves to develop a winning strategy. Notice that the Aumann's backward induction solution assumes common knowledge of rationality \cite{aum0,hal3}\footnote{Although according to Halpern, Stalnaker proved otherwise \cite{hal3,sta,sta0,sta1}.}. Granted, there can be several philosophical and epistemic issues about the centipede game and its relationship with rationality, but we will not pursue this direction here \cite{art6,art7}. 

This issue can also be approached from a dynamic epistemic perspective. Recently, it has been shown that in any game tree model $M$ taken as a PAL model, $\lim_{\mathsf{rational}} M$ is the actual subtree computed by the backward induction procedure where the proposition $\mathsf{rational}$ means that ``at the current node, no player has chosen a strictly dominated move in the past coming here'' \cite{ben14}. Therefore, the announcement of node-rationality produces the same result as the backward induction procedure. Each backward step in the backward induction procedure can then be obtained by the public announcement of node rationality. This result is quite impressive in the sense that it establishes a closer connection between communication and rationality, and furthermore leads to several more intriguing discussions about rationality. In this work, we refrain ourselves from pursuing this line of thought for the time being.

However, there seems to be a problem in topological models. The admissibility of limit models can take more than $\omega$ steps in topological models as we have conjectured earlier. Therefore, the BI procedure can take  $\omega$ steps or more. 

\begin{thm}
In topological models of games, under the assumption of rationality, the backward induction procedure can take more than $\omega$ steps. 
\end{thm}

\begin{proof}
Notice that each tree can easily be converted to a topology by taking the upward closed sets as opens. By the previous discussion, we know that backward induction solution can be attained by obtaining the limit models by publicly announcing the proposition $\mathsf{rationality}$. Therefore, by Theorem~\ref{baskent-limit}, stabilization can take more than $\omega$ step. Therefore, the corresponding backward induction scheme can also take more than $\omega$ step.
\end{proof}

This is indeed a problem about the attainability in infinite games: how can a player continue playing the game when she hit the limit ordinal $\omega$-th step in the backward induction procedure? In order not to diverge from our current focus, we leave this question open for further research.

\subsection{Persistence}

Let us now discuss stabilization in SSL framework. We already have a similar notion within the SSL context. Define \emph{persistent} formula in a model $M$ as the formula $\varphi$ whose truth is independent from the subsets in $M$. In other words, $\varphi$ is persistent if for all states $s$ and subsets $V \subseteq U$, we have $s, U \models \varphi$ impies $s, V \models \varphi$. Clearly, Boolean formulas are persistent in every model.

The significance of persistent formulas is the fact that they are independent of the subsets they occupy which means that they are immune to the epistemics of the model. Therefore, intuitively, they should also be immune to the \emph{changes} in the model. This is interesting due to the fact that now we have a quite strong way to tell what can and cannot be changed by public announcements in SSL.

\begin{thm}
Let $M$ be a model and $\varphi$ be persistent in $M$. Then, for any formula $\chi$ and neighborhood situation $(s, U)$, if $s, U \models \varphi$, then $s, U \models [\chi] \varphi$. In other words, true persistent formulas are immune to the public announcements.
\end{thm}

\begin{proof}
Proof follows directly from the definitions and the fact that after the public announcement of $\chi$, we always have $U_{\chi} \subseteq U$.
\end{proof}

In other words, we can have some formulas in SSL framework that are immune to the announcements.

\section{Conclusion and Future Work}

In this work, without using Kripke structures at all, we discussed PAL in two different geometrical systems. In subset space logic, we defined dynamic axioms for both epistemic and dynamic modalities and showed the corresponding completeness theorems. Moreover, we have applied the geometric ideas to model stabilization and persistent formulas. This gave us the connection regarding dynamic epistemic logics and rationality. We observed that in topological models, backward induction scheme loses its intuitiveness. There can be some mathematical solutions to this problem. For the backwards induction procedure that takes longer than $\omega$, modal-mu calculus can also be considered with its natural game theoretical semantics. Therefore, this can be a further research to see how $>\omega$-step backward induction scheme gets stabilized.

An interesting fact about topological models of modal logic is that only modal formulas can give an open or a closed set. However, one can stipulate that the extension of \emph{any} modal formula can be open or dually closed. If that is the case, one can obtain an incomplete or inconsistent logic respectively \cite{bas3,mor}. Moreover, some special algebras such as Heyting and co-Heyting algebras, correspond to those logics. Therefore, the topological investigation of PAL can be carried out in these special topological spaces or algebras in such a way that some special logics can further be investigated. This can also lead to the investigation of PAL in paraconsistent, paracomplete and dialethic frameworks where announcements and models may not be consistent nor complete.

\paragraph{Acknowledgement} This paper is an extended version of \cite{bas8,bas4}. We appreciate the feedback of FLAIRS-24 referees and Rohit Parikh.

\bibliographystyle{authordate3}  
\bibliography{/Users/can/Documents/Akademik/papers} 

\begin{thebibliography}{}

\bibitem[\protect\citename{Aiello {\em et~al.}, }2003]{aie2}
{\sc Aiello, Marco, van Benthem, Johan, \& Bezhanishvili, Guram}. 2003.
\newblock Reasoning About Space: the Modal Way.
\newblock {\em Journal of Logic and Computation}, {\bf 13}(6), 889--920.

\bibitem[\protect\citename{Artemov, }2009a]{art6}
{\sc Artemov, Sergei}. 2009a.
\newblock {\em Intelligent Players}.
\newblock Tech. rept. Department of Computer Science, The Graduate Center, The
  City University of New York.

\bibitem[\protect\citename{Artemov, }2009b]{art7}
{\sc Artemov, Sergei}. 2009b.
\newblock {\em Rational Decisions in Non-probablistic Setting}.
\newblock Tech. rept. TR-2009012. Department of Computer Science, The Graduate
  Center, The City University of New York.

\bibitem[\protect\citename{Aumann, }1995]{aum0}
{\sc Aumann, Robert~J.} 1995.
\newblock Backward Induction and Common Knowledge of Rationality.
\newblock {\em Games and Economic Behavior}, {\bf 8}(1), 6--19.

\bibitem[\protect\citename{Balbiani {\em et~al.}, }2007]{bal1}
{\sc Balbiani, Philippe, Baltag, Alexandru, van Ditmarsch, Hans, Herzig,
  Andreas, Hoshi, Tomohiro, \& de~Lima, Tiago}. 2007.
\newblock What Can We Achieve by Arbitrary Announcements? A Dynamic Take on
  Fitch's Knowability.
\newblock {\em In:} {\sc Samet, Dov} (ed), {\em Procedings of the 11th
  Conference on Theoretical Aspects of Rationality and Knowledge (TARK-2007)}.

\bibitem[\protect\citename{Balbiani {\em et~al.}, }2008]{bal3}
{\sc Balbiani, Philippe, Baltag, Alexandru, van Ditmarsch, Hans, Herzig,
  Andreas, \& de~Lima, Tiago}. 2008.
\newblock `{K}nowable' as `known after an announcement'.
\newblock {\em Review of Symbolic Logic}, {\bf 1}(3), 305--334.

\bibitem[\protect\citename{Baltag \& Moss, }2004]{bal2}
{\sc Baltag, Alexandru, \& Moss, Lawrence~S.} 2004.
\newblock Logics for Epistemic Programs.
\newblock {\em Synthese}, {\bf 139}(2), 165--224.

\bibitem[\protect\citename{Barwise, }1988]{bar}
{\sc Barwise, Jon}. 1988.
\newblock Three Views of Common Knowledge.
\newblock {\em Pages  365--379 of:} {\sc Publishers, Morgan~Kaufmann} (ed),
  {\em Proceedings of the 2nd conference on Theoretical aspects of reasoning
  about knowledge}.

\bibitem[\protect\citename{Ba{\c{s}}kent, }2007]{bas2}
{\sc Ba{\c{s}}kent, Can}. 2007 (July).
\newblock {\em Topics in Subset Space Logic}.
\newblock M.Phil. thesis, Institute for Logic, Language and Computation,
  Universiteit van Amsterdam.

\bibitem[\protect\citename{Ba{\c{s}}kent, }2011a]{bas8}
{\sc Ba{\c{s}}kent, Can}. 2011a.
\newblock Completeness of Public Announcement Logic in Topological Spaces.
\newblock {\em Bulletin of Symbolic Logic}, {\bf 17}(1), 142.

\bibitem[\protect\citename{Ba{\c{s}}kent, }2011b]{bas4}
{\sc Ba{\c{s}}kent, Can}. 2011b.
\newblock Geometric Public Announcement Logics.
\newblock
\newblock {\em Pages  87--88 of:} {\sc Murray, R.~Charles, \& McCarthy,
  Philip~M.} (eds), {\em Proceedings of the 24th Florida Artificial
  Intelligence Research Society Conference}AAAI Press, for FLAIRS-24.

\bibitem[\protect\citename{Ba{\c{s}}kent, }2011c]{bas3}
{\sc Ba{\c{s}}kent, Can}. 2011c.
\newblock Paraconsistency and Topological Semantics.
\newblock {\em http://arxiv.org/abs/1107.4939}.

\bibitem[\protect\citename{Bezhanishvili \& Gehrke, }2005]{bez2}
{\sc Bezhanishvili, Guram, \& Gehrke, Mai}. 2005.
\newblock Completeness of S4 with respect to the Real Line: Revisited.
\newblock {\em Annals of Pure and Applied Logic}, {\bf 131}(??), 287--301.

\bibitem[\protect\citename{Bezhanishvili {\em et~al.}, }2005]{bez1}
{\sc Bezhanishvili, Guram, Esakia, Leo, \& Gabelaia, David}. 2005.
\newblock Some Results on Modal Axiomatization and Definability for Topological
  Spaces.
\newblock {\em Studia Logica}, {\bf 81}(3), 325--55.

\bibitem[\protect\citename{Cate {\em et~al.}, }2009]{cat}
{\sc Cate, Balder~ten, Gabelaia, David, \& Sustretov, Dmitry}. 2009.
\newblock Modal Languages for Topology: Expressivity and Definability.
\newblock {\em Annals of Pure and Applied Logic}, {\bf 159}(1-2), 146--170.

\bibitem[\protect\citename{Fagin {\em et~al.}, }1995]{fag1}
{\sc Fagin, Ronald, Halpern, Joseph~Y., Moses, Yoram, \& Vardi, Moshe~Y.} 1995.
\newblock {\em Reasoning About Knowledge}.
\newblock MIT Press.

\bibitem[\protect\citename{Gabbay {\em et~al.}, }2003]{gabb}
{\sc Gabbay, D.~M., Kurucz, A., Wolter, F., \& Zakharyaschev, M.} 2003.
\newblock {\em Many Dimensional Modal Logics: Theory and Applications}.
\newblock Studies in Logic and the Foundations of Mathematics, vol. 145.
\newblock Elsevier.

\bibitem[\protect\citename{Goldblatt, }2006]{gold2}
{\sc Goldblatt, Robert}. 2006.
\newblock Mathematical Modal Logic: A View of Its Evolution.
\newblock {\em In:} {\sc Gabbay, Dov~M., \& Woods, John} (eds), {\em Handbook
  of History of Logic},  vol. 6.
\newblock Elsevier.

\bibitem[\protect\citename{Halpern, }2001]{hal3}
{\sc Halpern, Joseph~Y.} 2001.
\newblock Substantive Rationality and Backward Induction.
\newblock {\em Games and Economic Behavior}, {\bf 37}(2), 425--435.

\bibitem[\protect\citename{McKinsey \& Tarski, }1944]{mck1}
{\sc McKinsey, J. C.~C., \& Tarski, Alfred}. 1944.
\newblock The Algebra of Topology.
\newblock {\em The Annals of Mathematics}, {\bf 45}(1), 141--191.

\bibitem[\protect\citename{McKinsey \& Tarski, }1946]{mck}
{\sc McKinsey, J. C.~C., \& Tarski, Alfred}. 1946.
\newblock On Closed Elements in Closure Algebras.
\newblock {\em The Annals of Mathematics}, {\bf 47}(1), 122--162.

\bibitem[\protect\citename{Mortensen, }2000]{mor}
{\sc Mortensen, Chris}. 2000.
\newblock Topological Seperation Principles and Logical Theories.
\newblock {\em Synthese}, {\bf 125}(1-2), 169--178.

\bibitem[\protect\citename{Moss \& Parikh, }1992]{par9}
{\sc Moss, Lawrence~S., \& Parikh, Rohit}. 1992.
\newblock Topological Reasoning and the Logic of Knowledge.
\newblock {\em Pages  95--105 of:} {\sc Moses, Yoram} (ed), {\em Proceedings of
  TARK IV}.

\bibitem[\protect\citename{Parikh, }1991]{par10}
{\sc Parikh, Rohit}. 1991.
\newblock Finite and Infinite Dialogues.
\newblock {\em Pages  481--498 of:} {\sc Moschovakis, Y.} (ed), {\em
  Proceedings of a Workshop on Logic from Computer Science}.
\newblock Springer.

\bibitem[\protect\citename{Parikh {\em et~al.}, }2007]{par1}
{\sc Parikh, Rohit, Moss, Lawrence~S., \& Steinsvold, Chris}. 2007.
\newblock Topology and Epistemic Logic.
\newblock {\em In:} {\sc Aiello, Marco, Pratt-Hartman, Ian~E., \& van Benthem,
  Johan} (eds), {\em Handbook of Spatial Logics}.
\newblock Springer.

\bibitem[\protect\citename{Plaza, }1989]{pla}
{\sc Plaza, Jan~A.} 1989.
\newblock Logic of Public Communication.
\newblock {\em Pages  201--216 of:} {\sc Emrich, M.~L., Pfeifer, M.~S.,
  Hadzikadic, M., \& Ras, Z.~W.} (eds), {\em 4th International Symposium on
  Methodologies for Intelligent Systems}.

\bibitem[\protect\citename{Schwalbe \& Walker, }2001]{sch}
{\sc Schwalbe, Ulrich, \& Walker, Paul}. 2001.
\newblock Zermelo and the Early History of Game Theory.
\newblock {\em Games and Economic Behavior}, {\bf 34}(1), 123--137.

\bibitem[\protect\citename{Stalnaker, }1994]{sta0}
{\sc Stalnaker, Robert}. 1994.
\newblock On the Evaluation of Solution Concepts.
\newblock {\em Theory and Decision}, {\bf 37}(1), 49--73.

\bibitem[\protect\citename{Stalnaker, }1996]{sta1}
{\sc Stalnaker, Robert}. 1996.
\newblock Knowledge, Belief and Counterfactual Reasoning in Games.
\newblock {\em Economics and Philosophy}, {\bf 12}(2), 133--163.

\bibitem[\protect\citename{Stalnaker, }1998]{sta}
{\sc Stalnaker, Robert}. 1998.
\newblock Belief Revision in Games: Forward and Backward Induction.
\newblock {\em Mathematical Social Sciences}, {\bf 36}(1), 31--56.

\bibitem[\protect\citename{van Benthem, }2006]{ben10}
{\sc van Benthem, Johan}. 2006.
\newblock "One is a Lonely Number": Logic and Communication.
\newblock {\em In:} {\sc Chatzidakis, Z., Koepke, P., \& Pohlers, W.} (eds),
  {\em Logic Colloquium '02}.
\newblock Lecture Notes in Logic, vol. 27.
\newblock Association for Symbolic Logic.

\bibitem[\protect\citename{van Benthem, }2007]{ben18}
{\sc van Benthem, Johan}. 2007.
\newblock Rational Dynamics and Epistemic Logic in Games.
\newblock {\em International Game Theory Review}, {\bf 9}(1), 13--45.

\bibitem[\protect\citename{van Benthem \& Bezhanishvili, }2007]{ben2}
{\sc van Benthem, Johan, \& Bezhanishvili, Guram}. 2007.
\newblock Modal Logics of Space.
\newblock {\em In:} {\sc Aiello, Marco, Pratt-Hartman, Ian~E., \& van Benthem,
  Johan} (eds), {\em Handbook of Spatial Logics}.
\newblock Springer.

\bibitem[\protect\citename{van Benthem \& Gheerbrant, }2010]{ben14}
{\sc van Benthem, Johan, \& Gheerbrant, Amelie}. 2010.
\newblock Game Solution, Epistemic Dynamics and Fixed-Point Logics.
\newblock {\em Fundamenta Infomaticae}, {\bf 100}(1-4), 19--41.

\bibitem[\protect\citename{van Benthem \& Sarenac, }2004]{ben3}
{\sc van Benthem, Johan, \& Sarenac, Darko}. 2004.
\newblock The Geometry of Knowledge.
\newblock {\em Pages  1--31 of:} {\em Aspects of Universal Logic}.
\newblock Travaux Logic, vol. 17.

\bibitem[\protect\citename{van Benthem {\em et~al.}, }2005]{ben4}
{\sc van Benthem, Johan, van Eijck, Jan, \& Kooi, Barteld}. 2005.
\newblock {\em Logics of Communication and Change}.
\newblock Tech. rept. Institute for Logic, Language and Computation.

\bibitem[\protect\citename{van Benthem {\em et~al.}, }2006]{ben1}
{\sc van Benthem, Johan, Bezhanishvili, Guram, Cate, Balder~ten, \& Sarenac,
  Darko}. 2006.
\newblock Modal Logics for Product Topologies.
\newblock {\em Studia Logica}, {\bf 84}(3), 375--99.

\bibitem[\protect\citename{van Ditmarsch {\em et~al.}, }2007]{dit}
{\sc van Ditmarsch, Hans, van~der Hoek, Wiebe, \& Kooi, Barteld}. 2007.
\newblock {\em Dynamic Epistemic Logic}.
\newblock Springer.

\end{thebibliography}
\end{document}